\documentclass[showpacs,preprintnumbers,10pt,twocolumn]{revtex4}%
\usepackage{epsfig,float}
\usepackage{amssymb}
\usepackage{amsfonts}
\usepackage{amsmath}
\usepackage{graphicx}
\usepackage{dcolumn}
\usepackage{bm}
\usepackage{revsymb}%
\begin{document}
\title{Ratchet universality in the presence of thermal noise}
\author{Pedro J. Mart\'{\i}nez$^{1}$ and Ricardo Chac\'{o}n$^{2}$}
\affiliation{$^{1}$Departamento de F\'{\i}sica Aplicada, E.I.N.A., Universidad de Zaragoza,
E-50018 Zaragoza and Instituto de Ciencia de Materiales de Arag\'{o}n,
CSIC-Universidad de Zaragoza, E-50009 Zaragoza, Spain, EU }
\affiliation{$^{2}$Departamento de F\'{\i}sica Aplicada, Escuela de Ingenier\'{\i}as
Industriales, Universidad de Extremadura, Apartado Postal 382, E-06006
Badajoz, Spain, EU}
\date{\today}

\begin{abstract}
We show that directed ratchet transport of a driven overdamped Brownian
particle subjected to a spatially periodic and symmetric potential can be
reliably controlled by tailoring a biharmonic temporal force, in coherence
with the degree-of-symmetry-breaking mechanism. We demonstrate that the effect
of finite temperature on the purely deterministic ratchet scenario can be
understood as an \textit{effective noise-induced change} of the potential
barrier which is in turn controlled by the degree-of-symmetry-breaking
mechanism. Remarkably, we find that the same universal scenario holds for any
symmetric periodic potential, while optimal directed ratchet transport occurs
when the impulse transmitted (spatial integral over a half-period) by the
symmetric spatial force is maximum.

\end{abstract}

\maketitle

Directed transport without any net external force, the ratchet effect [1-3],
has been an intensely studied interdisciplinary subject over the last few
decades owing to its relevance in biology where ratchet mechanisms are found
to underlie the working principles of molecular motors [4,5], and to its wide
range of potential technological applications including micro- and
nano-technologies. Directed ratchet transport (DRT) is today understood to be
a result of the interplay of symmetry breaking [6], nonlinearity, and
non-equilibrium fluctuations, in which these fluctuations include temporal
noise [2], spatial disorder [7], and quenched temporal disorder [8]. In
extremely small systems, including many of those occurring in biological and
liquid environments as well as many nanoscale devices, DRT is often suitably
described by overdamped ratchets, in which inertial effects are negligible in
comparison with friction effects [2,9-11]. Here, we show how ratchet
universality [12] works subtly in the context of \textit{noisy} overdamped
ratchets by studying the dynamics of a universal model -- a Brownian particle
moving on a periodic substrate subjected to a biharmonic excitation [2,3],%
\begin{align}
\overset{.}{x}+\sin x  &  =\sqrt{\sigma}\xi\left(  t\right)  +\gamma
F_{bihar}\left(  t\right)  ,\nonumber\\
F_{bihar}\left(  t\right)   &  \equiv\eta\sin\left(  \omega t\right)
+2\left(  1-\eta\right)  \sin\left(  2\omega t+\varphi\right)  , \tag{1}%
\end{align}
where $\gamma$ is an amplitude factor, and the parameters $\eta\in\left[
0,1\right]  $ and $\varphi$ account for the relative amplitude and initial
phase difference of the two harmonics, respectively, while $\xi\left(
t\right)  $ is a Gaussian white noise with zero mean and $\left\langle
\xi\left(  t\right)  \xi\left(  t+s\right)  \right\rangle =\delta\left(
s\right)  $, and $\sigma=2k_{b}T$ with $k_{b}$ and $T$ being the Boltzmann
constant and temperature, respectively. For deterministic ratchets, this has
been shown to also be the case for topological solitons [8] and matter-wave
solitons [13]. It is worth noting that, in spite of the abundance of numerical
findings, the theoretical understanding of the directed transport phenomena
represented by Eq.~(1) remains far from being satisfactory [14] even about
half a century after the earliest studies [15-17]. Indeed, \textit{all} the
earlier theoretical predictions (cf. Refs. [3,6,16,17,18]) indicate that the
dependence of DRT velocity on the amplitudes of the two harmonics should scale
as
\begin{equation}
v\sim\gamma^{2}\eta^{2}\left(  1-\eta\right)  . \tag{2}%
\end{equation}
Note that this expression presents, as a function of $\eta$, a single maximum
at $\eta=2/3$, irrespective of the particular value of temperature, including
the limiting value $T=0$ (cf. Refs. [3,18]). The occurrence of DRT in Eq.~(1)
implies the breakage of two temporal symmetries: the shift symmetry and the
time-reversal symmetry of the biharmonic excitation [2]. For
\textit{deterministic} ratchets subjected to biharmonic forces, it has been
shown [12] that there exists a universal force waveform which optimally
enhances directed transport by symmetry breaking. This universal waveform is a
direct consequence of the degree-of-symmetry-breaking (DSB) mechanism. Indeed,
it has been shown that optimal enhancement of DRT is achieved when maximally
effective (i.e., \textit{critical}) symmetry breaking occurs (see [12] for
additional details). Since thermal noise is significant in magnitude and
unavoidable in diverse physical contexts, the following question naturally
arises: How does the DSB mechanism work at finite temperatures?

In this Letter, we shall address this important question and provide
analytical estimates for the dependence of the DRT on the system's parameters
which are in excellent agreement with numerical results. To study numerically
the effect of thermal noise $\left(  \sigma>0\right)  $ on the purely
deterministic ratchet scenario, we calculated the mean velocity on averaging
over different realizations of noise $\left\langle \left\langle \overset{.}%
{x}\right\rangle \right\rangle \equiv v$ (cf. Eq.~(1)). Since Gaussian white
noise does not break any relevant symmetry of Eq.~(1), and the ratchet
universality [12] predicts (for $\sigma=0$) that the optimal value of the
relative amplitude $\eta$ comes from the condition that the amplitude of
$\sin\left(  \omega t\right)  $ must be twice as large as that of $\sin\left(
2\omega t+\varphi\right)  $ in Eq. (1), na\"{\i}vely, one might expect that
the average velocity should present, as a function of $\eta$, a single maximum
at $\eta=\eta_{opt}\equiv4/5$ when $\varphi\neq0,\pi$, as in the purely
deterministic case. However, our numerical estimates of the $\eta$ value at
which the average velocity is maximum, $\eta_{opt}^{\sigma>0}$, indicate a
systematic deviation from $\eta_{opt}\equiv4/5$: $\Delta\eta\equiv
4/5-\eta_{opt}^{\sigma>0}>0$, which is independent of the noise intensity over
a significant finite range, as is shown in Figs. 1(a) and 2. 
\begin{figure}[htb]
\begin{center}
\epsfig{file=fig1a.eps,width=0.4\textwidth}
\epsfig{file=fig1b.eps,width=0.4\textwidth}
\epsfig{file=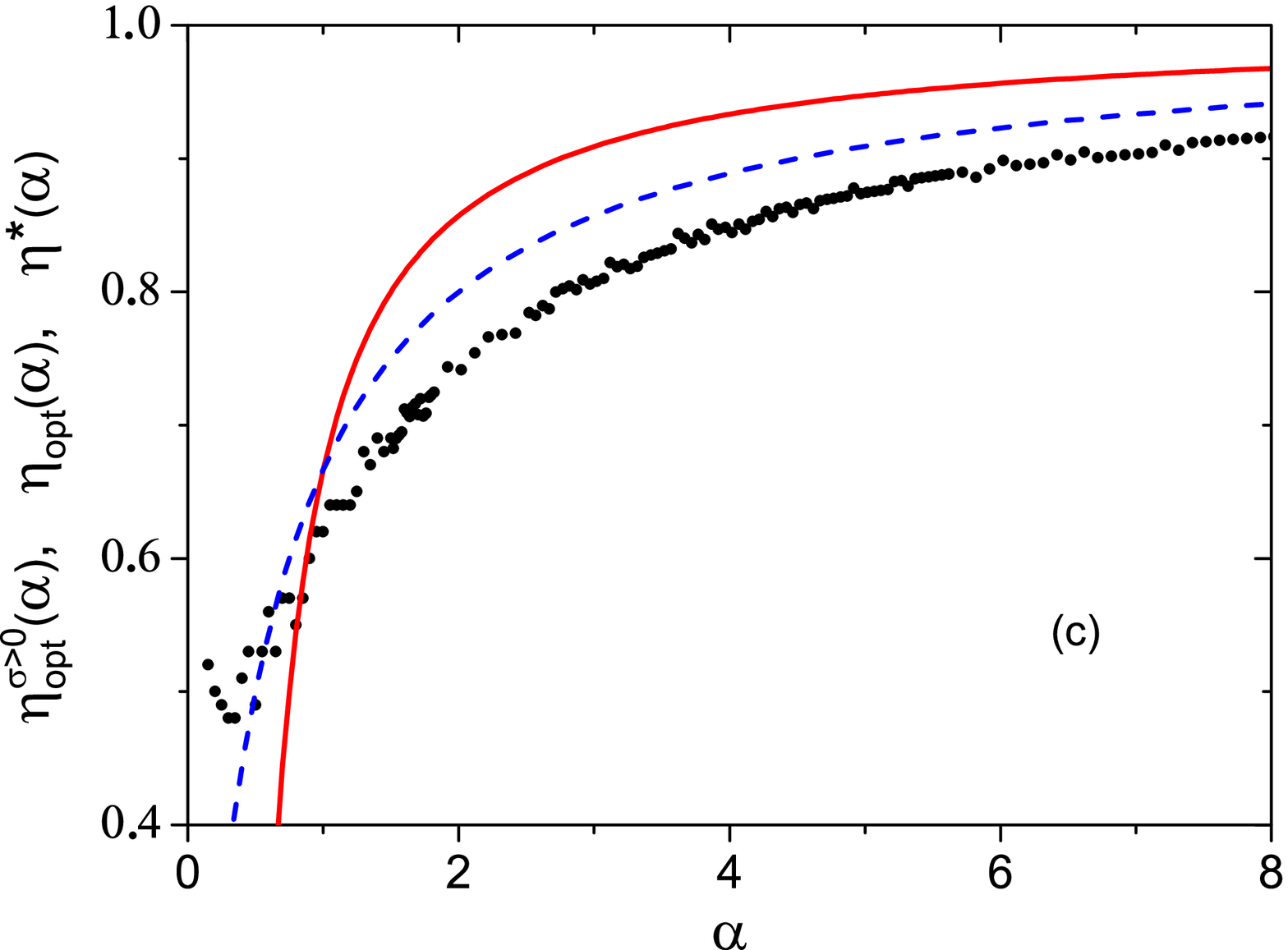,width=0.4\textwidth}
\end{center}
\vspace{-.4cm}
\caption{(Color online) (a) Average velocity $\left\langle \left\langle
\overset{.}{x}\right\rangle \right\rangle \equiv\left\langle \left\langle
\Lambda dx/dt^{\prime}\right\rangle \right\rangle $ [cf. Eqs.~(1) and (4)]
versus relative amplitude $\eta$ for $\varphi=\varphi_{opt}\equiv\pi
/2,\omega=0.08\pi,\gamma=2$, and three values of the noise intensity. (b)
Normalized biharmonic function [Eq. (3)] versus time for $\omega=1$ and three
values of $\eta$. (c) Value of $\eta$ where the average velocity is maximum,
$\eta_{opt}^{\sigma>0}$, versus $\alpha$ [cf. Eq.~(1) with $F_{bihar}^{\prime
}\left(  t\right)  $ instead of $F_{bihar}\left(  t\right)  $, see the text]
for $\varphi=\varphi_{opt}\equiv\pi/2,\omega=0.08\pi,\gamma=2$, and $\sigma
=1$. Also plotted is the theoretical prediction for the purely deterministic
case $\eta_{opt}\left(  \alpha\right)  \equiv2\alpha/\left(  1+2\alpha\right)
$ (dashed line) and the function $\eta^{\ast}(\alpha)$ (see the text, solid line).}
\label{fig_1}
\end{figure}
Note that these results are at variance with the prediction from Eq. (2) (i.e., $\eta
_{opt}\equiv2/3$). To explain this paradox, the following remarks are in
order. First, the effect of noise on the DRT strength depends on the amplitude
of the biharmonic excitation while keeping the remaining parameters constant.
Second, while changing $\eta$ and $\varphi$ allows one to control the breakage
of the aforementioned relevant symmetries, it also changes the amplitude of
the biharmonic excitation. Since the strength of any transport (whether or not
induced by symmetry breaking) depends on the amplitude of the driving
excitation, one concludes that these two effects of changing $\eta$ or
$\varphi$ overlap, so that one will find it difficult to distinguish the
contribution to transport that is purely due to symmetry breaking, and
\textit{hence} to clarify the interplay between noise and symmetry breaking.
We shall therefore consider an affine transformation of the biharmonic
excitation $F_{bihar}\left(  t\right)  $, for the optimal value $\varphi
=\varphi_{opt}\equiv\pi/2$ for example [19], to change its image to $\left[
-1/2,1/2\right]  ,\forall\eta$, thus making it possible to characterize the
genuine effect of noise on the purely deterministic ratchet scenario:%
\begin{equation}
f_{\varphi=\pi/2}^{\ast}\left(  t\right)  \equiv\frac{F_{bihar}\left(
t\right)  -m}{M-m}-\frac{1}{2},\tag{3}%
\end{equation}
where $m=m\left(  \eta\right)  \equiv\eta-2,\forall\eta$, while $M=M\left(
\eta\right)  \equiv\frac{\eta^{2}+32\left(  1-\eta\right)  ^{2}}{16\left(
1-\eta\right)  }$ for $0\leqslant\eta\leqslant8/9$ and $M\left(  \eta\right)
\equiv3\eta-2$ for $8/9\leqslant\eta\leqslant1$ (see Fig.~1(b)). After
substituting Eq.~(3) into Eq.~(1), one straightforwardly obtains%
\begin{equation}
\frac{dx}{dt^{\prime}}+\frac{1}{\Lambda}\sin x=\gamma W+\gamma f_{\varphi
=\pi/2}^{\ast}\left(  t^{\prime}\right)  +\sqrt{\sigma^{\prime}}\xi\left(
t^{\prime}\right)  ,\tag{4}%
\end{equation}
where $\Lambda=\Lambda\left(  \eta\right)  \equiv M-m,t^{\prime}=t^{\prime
}\left(  t,\eta\right)  \equiv\Lambda t,W=W\left(  \eta\right)  \equiv\left(
M+m\right)  /\left(  2\Lambda\right)  $, $\omega^{\prime}=\omega^{\prime
}\left(  \omega,\eta\right)  \equiv\omega/\Lambda$, and $\sigma^{\prime
}=\sigma^{\prime}\left(  \sigma,\eta\right)  \equiv\sigma/\Lambda$. It is
worth noting that the function $\Lambda\left(  \eta\right)  $ is merely the
width of the image of $F_{bihar}\left(  t\right)  $, i.e., the difference
between its maxima and minima as a function of $\eta$, and that it presents a
single minimum at $\eta=\eta^{\ast}\equiv6/7$. Also, the function $W\left(
\eta\right)  $ represents an $\eta$-dependent \textquotedblleft
load\textquotedblright\ force having a single maximum (in absolute value) at
$\eta=\eta_{opt}\equiv4/5$, while $W\left(  \eta=0,1\right)  =0$, as expected.
These particular values of $\eta^{\ast}$ and $\eta_{opt}$ are a direct
consequence of the application of ratchet universality to the specific form of
the present biharmonic excitation $F_{bihar}\left(  t\right)  $. However, to
better understand the roots of the present problem, it is convenient to
consider the more general form $F_{bihar}^{\prime}\left(  t\right)  \equiv
\eta\sin\left(  \omega t\right)  +\alpha\left(  1-\eta\right)  \sin\left(
2\omega t+\varphi\right)  $, with $\alpha>0$ being a parameter. For this case,
one has $\eta^{\ast}=\eta^{\ast}(\alpha),\eta_{opt}=\eta_{opt}\left(
\alpha\right)  \equiv2\alpha/\left(  1+2\alpha\right)  $, while the deviation
$\Delta\eta\left(  \alpha\right)  \equiv\eta_{opt}\left(  \alpha\right)
-\eta_{opt}^{\sigma>0}\left(  \alpha\right)  $ suggests a certain correlation
between $\eta_{opt}\left(  \alpha\right)  $ and $\eta_{opt}^{\sigma>0}\left(
\alpha\right)  $ over a wide range of $\alpha$ values from $\alpha\simeq2$
(see Fig.~1(c)). Once again, all the earlier theoretical analysis (cf. Refs.
[3,6,16,17,18]) predict (for $F_{bihar}^{\prime}\left(  t\right)  $) that the
average velocity should present, as a function of $\eta$, a single maximum at
$\eta=2/3$, irrespective of the particular values of temperature (including
the limiting value $T=0$) and parameter $\alpha$. Together, these results
therefore allow one to draw the following conclusions from Eq.~(4).

First, the aforementioned twofold transport effect of a biharmonic excitation
$F_{bihar}\left(  t\right)  $, as $\eta$ varies from $0$ to $1$, may be
decoupled into two terms: a constant excitation, $W$, and a biharmonic
excitation, $f_{\varphi=\pi/2}^{\ast}\left(  t^{\prime}\right)  $, having an
amplitude which is \textit{independent} of $\eta$. The relevant observation is
that both excitations yield a maximum strength of transport at $\eta
=\eta_{opt}\equiv4/5$. Therefore, replacing $F_{bihar}\left(  t\right)  $ with
$f_{\varphi=\pi/2}^{\ast}\left(  t\right)  $ in Eq.~(1),%
\begin{equation}
\overset{.}{x}+\sin x=\sqrt{\sigma}\xi\left(  t\right)  +\gamma f_{\varphi
=\pi/2}^{\ast}\left(  t\right)  , \tag{5}%
\end{equation}
should yield a maximum average velocity at $\eta=\eta_{opt}\equiv4/5$, as is
indeed confirmed by numerical experiments (see Fig.~3). Thus, we propose for
the system (5) the following scaling for the average velocity:%
\begin{equation}
\left\langle \left\langle \overset{.}{x}\right\rangle \right\rangle \sim
CW\left(  \eta\right)  , \tag{6}%
\end{equation}
where $C$ is a fitting constant that depends on the remaining system
parameters. Furthermore, for sufficiently high temperature (i.e., sufficiently
far from the \textquotedblleft steps\textquotedblright\ regime occurring at
$\sigma=0$, see Fig.~2) and driving amplitude (i.e., in the absence of
stochastic resonance effects), \textit{exact} agreement between numerical
results and scaling (6) is expected over the complete range of $\eta$ values
(see Fig.~3), while the scaling (6) remains valid over a wide range of
frequencies (data not shown).
\begin{figure}[htb]
\begin{center}
\epsfig{file=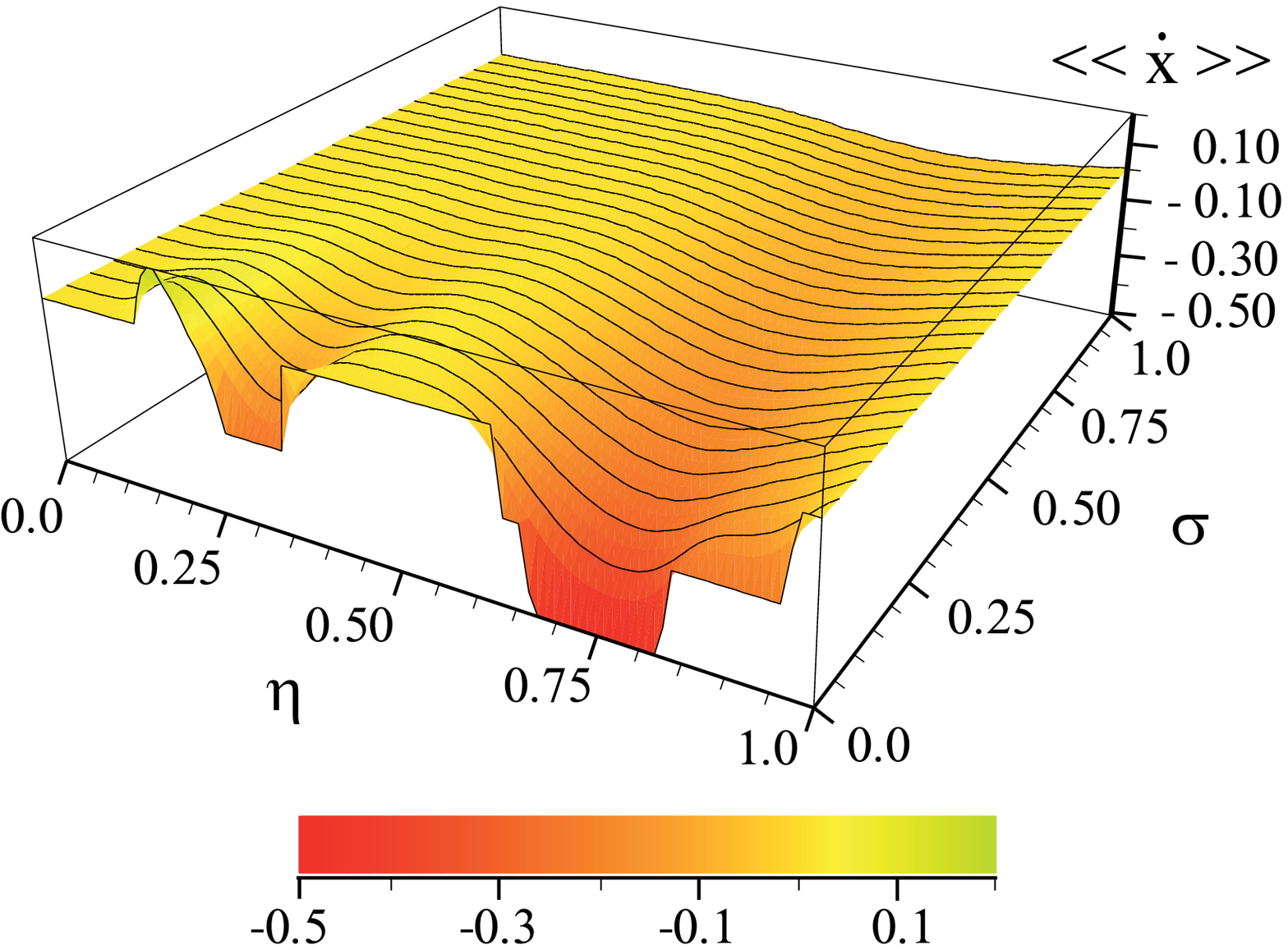,width=0.4\textwidth}
\epsfig{file=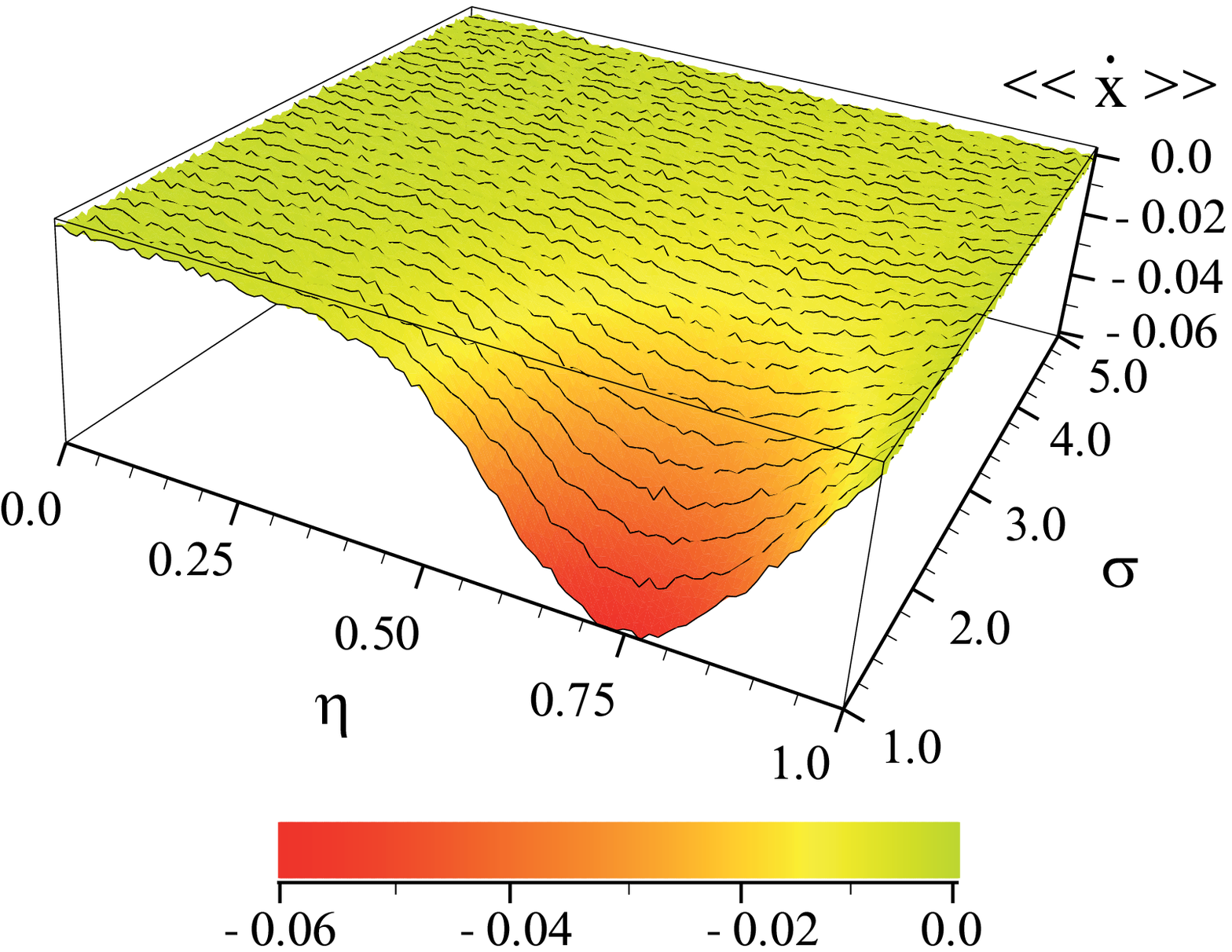,width=0.4\textwidth}
\end{center}
\vspace{-0.4cm}
\caption{(Color online) Average velocity $\left\langle \left\langle \overset
{.}{x}\right\rangle \right\rangle $ [cf. Eq.~(1)] versus relative amplitude
$\eta$ and noise intensity $\sigma$ for $\varphi=\varphi_{opt}\equiv\pi/2,$
$\omega=0.08\pi$, and $\gamma=2$. The ranges of small and large noise
intensities are shown in the top and bottom panels, respectively.}
\label{fig_2}
\end{figure}

\begin{figure}[htb]
\begin{center}
\epsfig{file=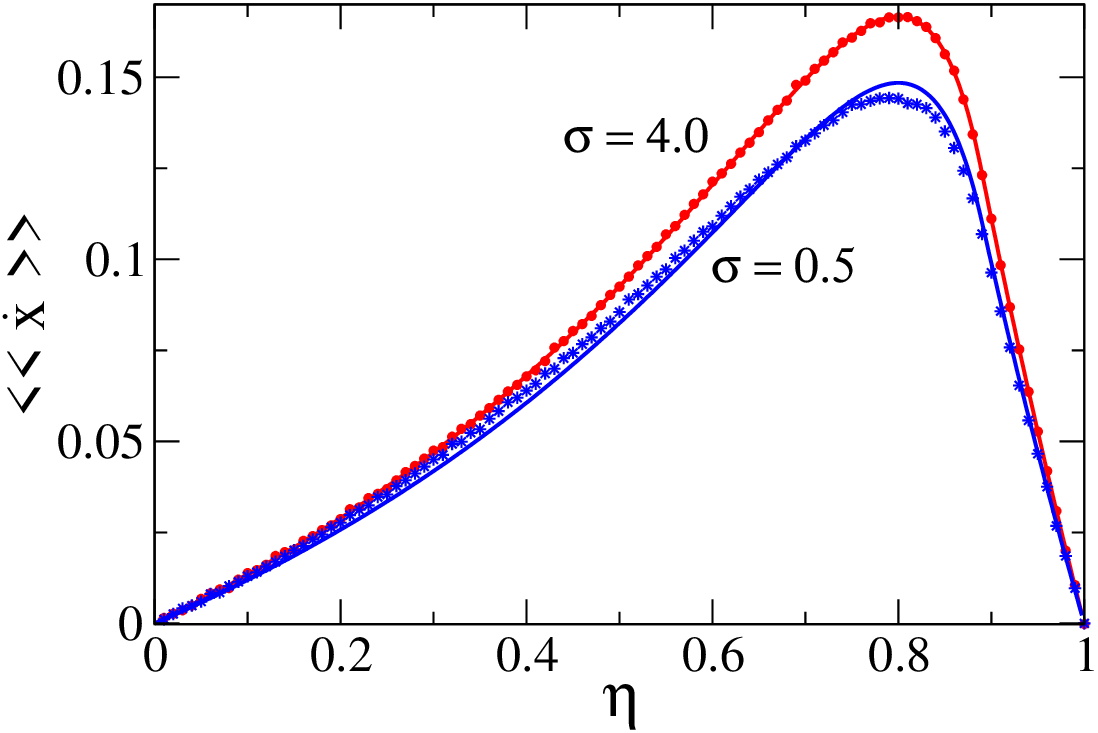,width=0.4\textwidth}
\end{center}
\vspace{-0.6cm}
\caption{(Color online) Average velocity $\left\langle \left\langle \overset
{.}{x}\right\rangle \right\rangle $ [cf. Eq.~(5)] versus relative amplitude
$\eta$ for $\omega=0.08\pi,\gamma=2,$ and two values of the noise intensity.
Also plotted is the scaling law (6) for $C=1.00017$ $\left(  \sigma=4\right)
$ and $C=0.89072$ $\left(  \sigma=0.5\right)  $ [cf. Eq.~(6); solid
lines].}
\label{fig_3}
\end{figure}

And second, alternatively to the case discussed in the first conclusion,
\textit{only} rescaling the temperature with the width $\Lambda$ in Eq.~(1) in
accordance with Eq.~(4),%
\begin{equation}
\overset{.}{x}+\sin x=\sqrt{\Lambda\sigma}\xi\left(  t\right)  +\gamma
F_{bihar}\left(  t\right)  ,\tag{7}%
\end{equation}
should also yield a maximum average velocity at $\eta=\eta_{opt}\equiv4/5$.
Numerical experiments also confirmed this prediction, as in the illustrative
examples shown in Fig.~4(a). Thus, it is only after rescaling $\sigma
\rightarrow\Lambda\sigma$ in Eq.~(1) that one recovers the purely
deterministic ratchet scenario, which was an unanticipated result. To make the
comparison between Eqs.~(1) and (7) clearer, let us first transform Eq.~(7)
into the equation%
\begin{align}
\overset{.}{x}+\frac{1}{\Lambda}\sin x &  =\sqrt{\sigma}\xi\left(  t\right)
\tag{8}\\
&  +\gamma^{\prime}\left[  \eta\sin\left(  \omega^{\prime}t\right)  +2\left(
1-\eta\right)  \sin\left(  2\omega^{\prime}t+\varphi\right)  \right]
\nonumber
\end{align}
by rescaling the time, $t\rightarrow\Lambda t$, and where $\gamma^{\prime
}\equiv\gamma/\Lambda,\omega^{\prime}\equiv\omega/\Lambda$. Since the
transport properties of Eq.~(8) are similar to those of Eq.~(7) in the sense
that their respective average velocities present a single extremum at the same
value of $\eta$ for a fixed set of the remaining parameters, and that such an
optimal value, $\eta_{opt}\equiv4/5$, is independent of the particular values
of the amplitude and the frequency of the biharmonic excitation [12], one
concludes from the comparison of Eqs.~(1) and (8) that the effect of thermal
noise on the purely deterministic ratchet scenario can be understood as an
\textit{effective noise-induced change} of the potential barrier $\left(
d=d\left(  \eta\right)  \equiv\Lambda\left(  \eta\right)  \right)  $ which is
in turn controlled by the degree-of-symmetry-breaking mechanism through the
function $\Lambda\left(  \eta\right)  $. Recall that the average velocity
$\left\langle \left\langle \overset{.}{x}\right\rangle \right\rangle $
exhibits, as a function of the potential barrier $d$, a single maximum due to
the thermal interwell activation (TIA) mechanism and the limiting behaviours
$\lim_{d\rightarrow0,\infty}\left\langle \left\langle \overset{.}%
{x}\right\rangle \right\rangle =0$ [14]. Obviously, the same scenario also
holds when $F_{bihar}\left(  t\right)  $ is replaced with $F_{bihar}^{\prime
}\left(  t\right)  $, which allows one to understand the behaviour of the
deviation $\Delta\eta\left(  \alpha\right)  \equiv\eta_{opt}\left(
\alpha\right)  -\eta_{opt}^{\sigma>0}\left(  \alpha\right)  $ as $\alpha$ is
changed (cf. Fig.~1(c)). Clearly, one can distinguish three \textit{regimes}.
Over the range $0<\alpha\lesssim1/2$, the TIA mechanism dominates over the DSB
mechanism. Indeed, the contribution of the DSB mechanism to directed transport
becomes ever smaller as $\alpha\rightarrow0$ because $\eta_{opt}\left(
\alpha\rightarrow0\right)  \rightarrow0$, and hence the (whole) amplitude of
the biharmonic excitation for which the DRT strength is maximum also becomes
ever smaller as $\alpha\rightarrow0$, while the contribution of the TIA
mechanism remains significant over the entire range $0<\alpha\lesssim1/2$.
Then one observes a transition regime over the range $1/2\lesssim
\alpha\lesssim2$ as the effect of the DSB mechanism strengthens, which is
manifest in the existence of a narrow range of $\alpha$ values in which
$\Delta\eta\left(  \alpha\right)  \approx0$ (see Fig.~1(c)). Finally, for the
range $\alpha\gtrsim2$, the contributions of the DSB and TIA mechanisms are
correlated in the sense of the aforementioned effective noise-induced change
of the potential barrier. 
\begin{figure}[htb]
\begin{center}
\epsfig{file=fig4a.eps,width=0.4\textwidth}
\vspace{.3cm}
\epsfig{file=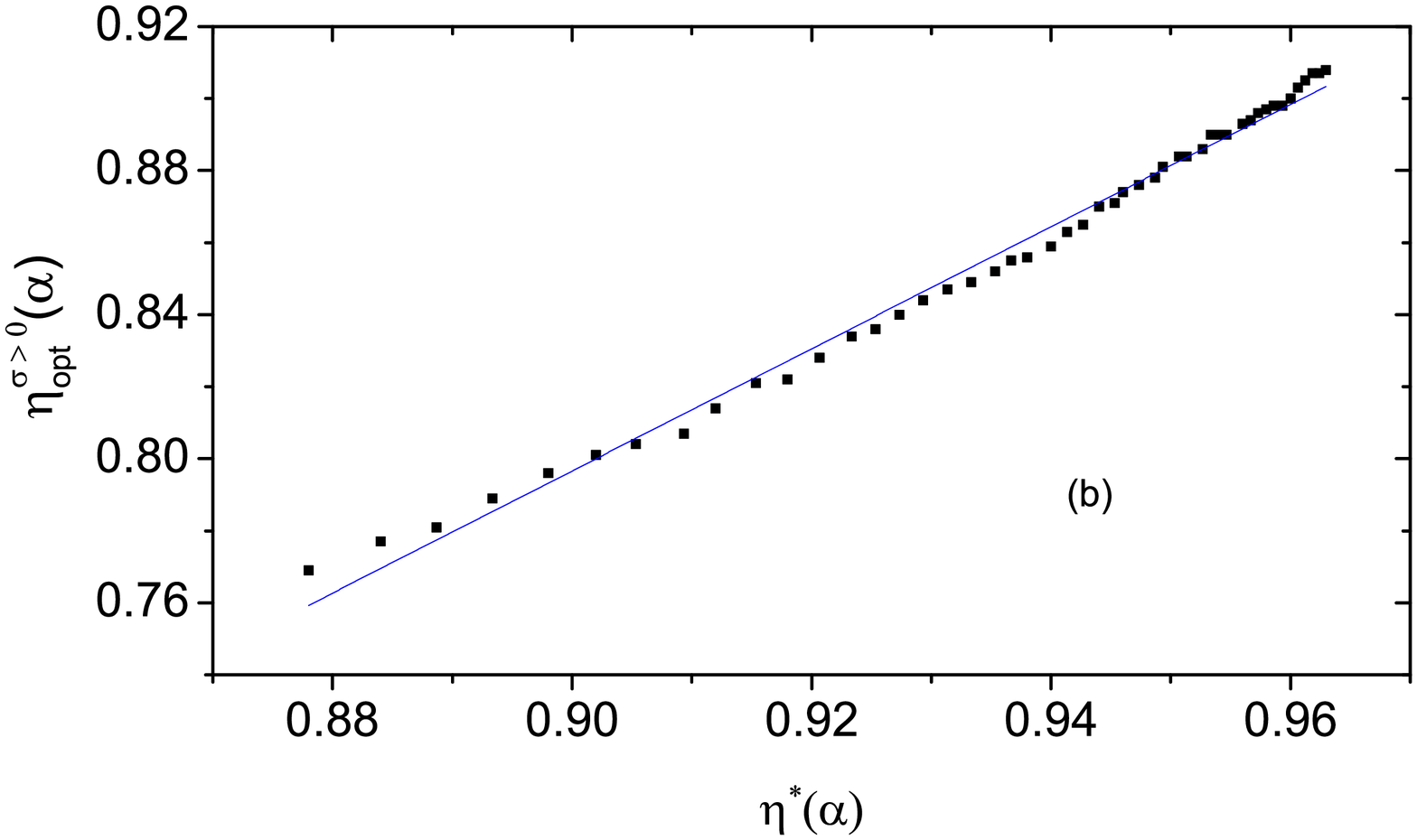,width=0.4\textwidth}
\vspace{.3cm}
\epsfig{file=fig4c.eps,width=0.4\textwidth}
\end{center}
\vspace{-0.6cm}
\caption{(Color online) (a) Average velocity $\left\langle \left\langle
\overset{.}{x}\right\rangle \right\rangle $ [cf. Eq.~(7)] versus relative
amplitude $\eta$ for $\omega=0.08\pi,\gamma=2$, and three values of the noise
intensity. (b) Value of $\eta$ where the average velocity is maximum,
$\eta_{opt}^{\sigma>0}\left(  \alpha\right)  $, versus function $\eta^{\ast
}(\alpha)$ [cf. Eq.~(1) with $F_{bihar}^{\prime}\left(  t\right)  $ instead of
$F_{bihar}\left(  t\right)  $, see the text] and linear fit (9) (solid line)
over the range $2.3\leqslant\alpha\leqslant7$ for $\varphi=\varphi_{opt}%
\equiv\pi/2,\omega=0.08\pi,\gamma=2$, and $\sigma=1$. (c) Average velocity
$\left\langle \left\langle \overset{.}{x}\right\rangle \right\rangle $ [cf.
Eq.~(1)] versus initial phase difference $\varphi$ for $\omega=0.08\pi
,\gamma=2,\sigma=1$, and three values of $\eta$.}
\label{fig_4}
\end{figure}
This means that $\eta_{opt}^{\sigma>0}\left(
\alpha\right)  $ is expected to be proportional to the value of the relative
amplitude where the effective noise-induced potential barrier presents a
minimum:%
\begin{equation}
\eta_{opt}^{\sigma>0}\left(  \alpha\right)  \sim A\eta^{\ast}(\alpha
)+B,\tag{9}%
\end{equation}
with $A\simeq1.71,B\simeq-0.73$ being constants that are \textit{independent}
of the remaining system parameters. The general scaling (9) is confirmed by
numerical experiments (see Fig.~4(b)). Also, with regard to the dependence of
the DRT strength on the initial phase difference, numerical results confirmed
the scaling $\left\langle \left\langle \overset{.}{x}\right\rangle
\right\rangle \sim C^{\prime}\sin\varphi$, where $C^{\prime}$ is a fitting
constant that depends on the remaining system parameters, in accordance with
ratchet universality [12] (see Fig.~4(c)). 
\begin{figure}[htb]
\begin{center}
\epsfig{file=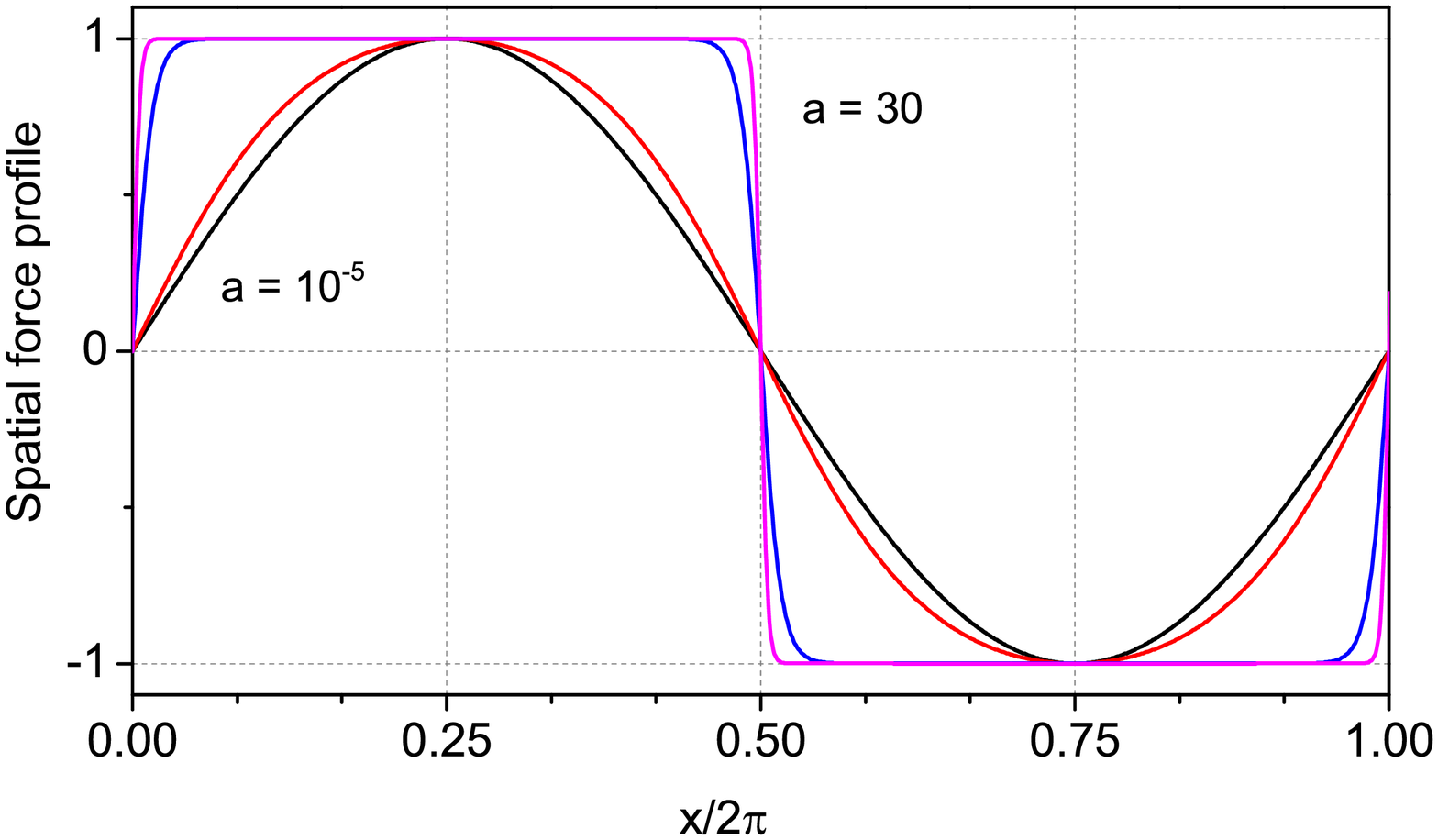,width=0.4\textwidth}
\epsfig{file=fig5b.eps,width=0.4\textwidth}
\end{center}
\vspace{-0.5cm}
\caption{(Color online) Top: Wave forms of the spatial force $-\tanh\left(
a\sin x\right)  /\tanh\left(  a\right)  $ for four values of the shape
parameter: $a=10^{-5}$ (nearly sinusoidal pulse), $a=1,a=10$, and $a=30$
(nearly square-wave pulse). Bottom: Average velocity $\left\langle
\left\langle \overset{.}{x}\right\rangle \right\rangle $ [cf. Eq.~(10)] versus
relative amplitude $\eta$ for $\varphi=\varphi_{opt}\equiv\pi/2,\omega
=0.08\pi,\gamma=2,\sigma=1$, and the same four values of the shape parameter
$a$.}
\label{fig_5}
\end{figure}

Finally, it is worth noting that
the same scenario holds when the present sinusoidal potential $V(x)\equiv-\cos
x$ (cf. Eq. (1)) is replaced with any symmetric (under reflection) periodic
potential. For the sake of clarity, we show this property by considering the
generalized model%
\begin{equation}
\overset{.}{x}+dU(x;a)/dx=\sqrt{\sigma}\xi\left(  t\right)  +\gamma
F_{bihar}\left(  t\right)  ,\tag{10}%
\end{equation}
where the spatial force $-dU(x;a)/dx\equiv-\tanh\left(  a\sin x\right)
/\tanh\left(  a\right)  $ has the same amplitude for any wave form (i.e.,
$\forall a\in\left[  0,\infty\right]  $). One has $\left[  dU(x;a)/dx\right]
_{a=0}=\sin x$ while, in the other limit, $\left[  dU(x;a)/dx\right]
_{a=\infty}$ is the square-wave function (see Fig. 5, top panel). We find that
optimum DRT occurs when the impulse transmitted (spatial integral over a
half-period) by the spatial force is maximum while keeping the remaining
parameters fixed, as is shown in Fig. 5 (bottom panel). Remarkably, the
general scaling (9) holds for any value of the shape parameter $a$ (see Fig.
6), thus confirming its universality [20].
\begin{figure}[htb]
\begin{center}
\epsfig{file=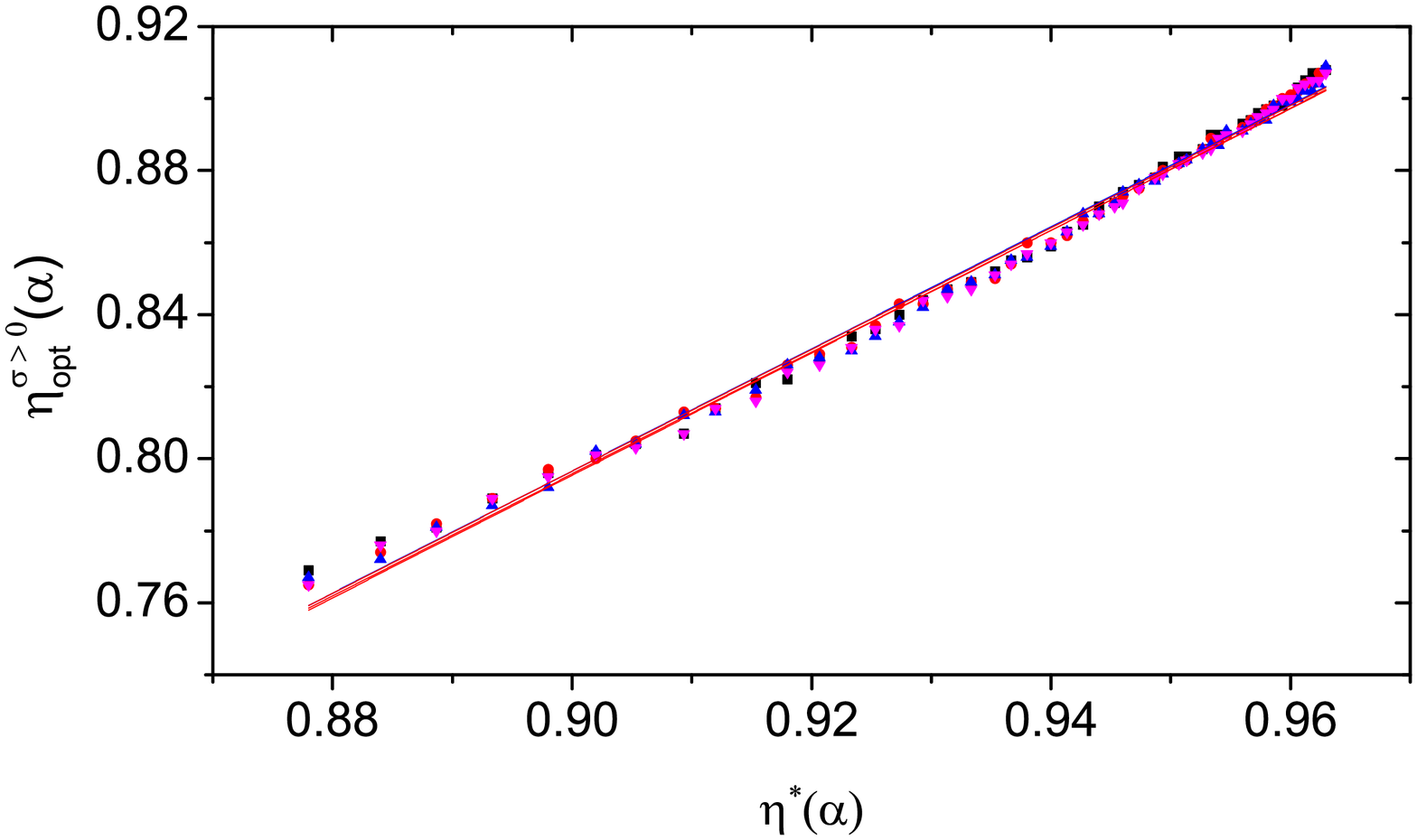,width=0.45\textwidth}
\end{center}
\vspace{-.3cm}
\caption{(Color online) Value of $\eta$ where the average velocity is maximum,
$\eta_{opt}^{\sigma>0}\left(  \alpha\right)  $, versus function $\eta^{\ast
}(\alpha)$ [cf. Eq.~(10) with $F_{bihar}^{\prime}\left(  t\right)  $ instead
of $F_{bihar}\left(  t\right)  $, see the text] and linear fit (9) (solid
line) over the range $2.3\leqslant\alpha\leqslant7$ for $\varphi=\varphi
_{opt}\equiv\pi/2,\omega=0.08\pi,\gamma=2,\sigma=1$, and four values of the
shape parameter: $a=10^{-5}$ (circles), $a=1$ (squares), $a=10$ (triangles),
and $a=30$ (stars).}
\label{fig_6}
\end{figure}

In summary, we have explained the interplay between thermal noise and symmetry
breaking in the ratchet transport of a Brownian particle moving on a periodic
substrate subjected to a temporal biharmonic excitation, in coherence with the
degree-of-symmetry-breaking mechanism. For any finite value of temperature,
including the deterministic limit $\left(  T=0\right)  $, the reason of the
failure of all the earlier theoretical predictions (cf. Refs. [3,6,16,17,18],
Eq. (2)) is now clear: both the moment expansion method [3,6,15] and the
functional Taylor expansion formalism [18] assume that the contributions of
the amplitudes of the two harmonics to the average velocity are
\textit{independent}. However, the existence of a universal waveform which
optimally enhances directed ratchet transport [12] implies that the two
amplitudes are correlated in the sense mentioned above. Thus, the general
ratchet scenario presented in this Letter provides a reliable theoretical
framework for the optimal control of the dynamics of Brownian ratchets in
future applications, including dimers and more complex systems.

P.J.M. acknowledges financial support from the Ministerio de Ciencia e
Innovaci\'{o}n (MICINN, Spain) through project FIS2011-25167 cofinanced by
FEDER funds. R.C. acknowledges financial support from the MICINN, Spain,
through project FIS2012-34902 cofinanced by FEDER funds, and from the Junta de
Extremadura (JEx, Spain) through project GR10045.


\begin{thebibliography}{99}
\bibitem {19}The analysis is quite similar for any other value of the initial
phase difference $\varphi$. In particular, for the other optimal value,
$\varphi_{opt}\equiv3\pi/2$, one has DRT with the same strength but opposite
direction to that corresponding to $\varphi_{opt}\equiv\pi/2$ for each value
of $\eta\in\left]  0,1\right[  $ (cf. Ref. [12]).
\bibitem {20}The same excellent agreement with scaling (9) was found for other
alternative symmetric spatial potentials (data not shown).
\end{thebibliography}
\end{document}